\documentclass{aastex} 
\usepackage[twocolumn]{emulateapj5}

\def\sameauthor{\underbar{\qquad\qquad}.}
\def\halpha{H$\alpha$}
\def\hbeta{H$\beta$}
\def\hgamma{H$\gamma$}

\slugcomment {Submitted to ApJ, Part 1}

\shortauthors{SCHMIDT \& SMITH}
\shorttitle{RADIO-OPTICAL ALIGNED QUASARS}

\begin{document}

\title{Evidence for Polarized Synchrotron Components \\
in Radio-Optical Aligned Quasars}

\author{\centerline{Gary D. Schmidt}}

\affil{Steward Observatory, The University of Arizona, Tucson, AZ 85721}

\email{gschmidt@as.arizona.edu}

\and

\author{\centerline{Paul S. Smith\altaffilmark{1}}}

\altaffiltext{1}{Current address: Steward Observatory, The
University of Arizona, Tucson, AZ 85721, psmith@as.arizona.edu}

\affil{National Optical Astronomy Observatories, Kitt Peak National
Observatory, Box 26732, Tucson, AZ 85726}

\begin{abstract}

A tendency for the axes of double-lobed radio quasars to be aligned with the
electric vectors of optical polarization in the active galactic nuclei is
well-known.  However, the origin of the polarization and reason behind its
correlation with radio morphology is not yet established.  From accurate
spectropolarimetry of 7 quasars which show this alignment effect, we find that
the polarization is universally confined to the continuum, and not shared by
the line emission or 3000~\AA\ bump.  Over the observed region
$4000-8000$~\AA, the spectral indices of polarized flux, $P\times F_\lambda$,
are consistent with uniform polarization applied to the light of the big blue
bump or with synchrotron emission. However, electron scattering from an
optically-thick, geometrically-thin, accretion disk is well-known to give rise
to a polarization position angle that is perpendicular to the disk axis.
Optical synchrotron emission akin to that shown to exist in the miniblazar
3C~273 is a more attractive explanation, and supporting evidence can be found
for some of the targets in the form of polarimetric variability over intervals
of years. Properties of the most strongly-polarized radio-optical aligned
quasars can be explained by misdirected blazar core components that have net
polarizations of $P\sim10$\% and provide $\sim$10\% of the total optical light.

\end{abstract}


\section{Introduction}

The ``Unified Scheme'' for active galactic nuclei (AGN) has enjoyed remarkable
success in explaining not only the spectroscopic and polarimetric properties
of various classes of objects, but also their morphologies in the light of
certain emission lines, the overall radio structure and spectra of quasars,
and the existence of IR-luminous, but optically-faint AGN (e.g. Antonucci
1993; Cimatti et al. 1993; Capetti, Macchetto, \& Lattanzi 1997; Hines et al.
1995). The idea relies on the presence of a nominally toroidal structure
surrounding the central engine which obscures the compact object and
broad-line region (BLR) for high-inclination sightlines. From such vantage
points, the bright nuclear emission components are seen primarily in light
which has become polarized by scattering off dust and/or electrons located
above and below the polar openings of the torus.  For these objects (e.g.
Seyfert 2 galaxies, narrow-line radio galaxies, {\it IRAS\/} QSOs), extinction
of the direct optical and near-IR continuum also increases the contrast of the
emission from the more ubiquitous narrow-line region (NLR), earning their
type-2 spectroscopic designation and resulting in extreme optical/far-IR
colors.

By contrast, objects of type 1 are thought to be viewed through the toroidal
hole, a perspective which provides the observer with a relatively unattenuated
view of the central engine and BLR. Type-1 objects include the broad-line
quasars, broad-line radio galaxies, and Seyfert 1 galaxies. When our sightline
lies very near the axis of a relativistic jet emanating from the compact
object, Doppler-boosted synchrotron emission from the jet dominates the
spectrum and a highly variable and polarized blazar is observed.  Though the
model does not explain the essential difference between radio-loud and -quiet
AGN, it successfully applies within each radio class.

Stockman, Angel, \& Miley (1979) were the first to point out that the
$E$-vector of optical polarization for quasars associated with double-lobe
radio sources tends to be aligned with the radio source axis on the sky.  This
tendency toward ``parallel'' polarization was soon found to exist also among
Seyfert 1 galaxies (Antonucci 1983) and radio galaxies (Antonucci 1982, 1984;
Rudy et al. 1983), and to extend to radio structure at the milliarcsecond
scale (Rusk \& Seaquist 1985). Though interest in the radio-optical alignment
was soon eclipsed by Antonucci's discovery (1983; see also Antonucci \& Miller
1985) of very strong perpendicular polarization in some type-2 Seyfert
galaxies, the result of Stockman et al. (1979) was early evidence for
aspect-dependent classification of AGN.

There is no doubt that the dominant polarizing mechanism for most type-2
objects is small-particle scattering.  Spectra of polarized flux ($P\times
F_\lambda$) often show broad permitted lines that are not evident in the total
flux spectrum but are characteristic of a classical emission-line QSO.  In
addition, the polarized continuum occasionally exhibits the ``3000~\AA\
bump'', a pseudo-continuum from \ion{Fe}{2} and Balmer free-bound emission
arising within the BLR (Grandi 1981; Wills, Netzer, \& Wills 1985). The
perpendicularity between the optical polarization $E$-vector and radio
morphology is simply a result of the scatterers lying along the same general
direction from the central source as the plasmons. In contrast, the
polarization properties of radio-loud, non-blazar, type-1 AGN are not
well-studied. Overall, the polarization of these objects rarely exceeds 2\%
(Stockman, Moore, \& Angel 1984), so the sources are difficult to study in
detail. The interpretation of weak polarization is also more ambiguous, since
synchrotron emission, scattering in an extended halo or accretion disk, or
even partial extinction by a medium of aligned dust grains are all potentially
viable mechanisms (e.g., Stockman et al. 1979; Antonucci 1984). Discrimination
among these possibilities requires spectropolarimetry of sufficient quality to
measure the polarization of the emission lines {\it vs.\/} the continuum and to
compare the continuum shape in polarized flux with that in total flux. Such a
study is justified by the potential for new, qualitative information on the
structure of the inner regions of QSOs and possibly on the salient distinction
between radio-loud and -quiet objects.

\section{Observations and Object Selection}

In this paper we present the results of linear spectropolarimetry of 8
double-lobed quasars from the original Stockman et al. (1979) list which show
alignment between the axis of radio emission and optical polarization.  The
data were obtained using the Steward Observatory 2.3~m Bok telescope and CCD
Spectropolarimeter (Schmidt, Stockman, \& Smith 1992).  Wavelength coverage
generally spans a single order of the low-resolution grating,
$\lambda\lambda4000-8000$, however on occasion a second tilt was used to
include an important emission line.  The data typically consist of $10-20$
observations from several epochs over the period $1995-1999$, each observation
being the result of $\sim$20~min of integration in a $Q-U$ sequence of
waveplate orientations.  Entrance slits were $3\arcsec-4\arcsec$ width
$\times$ 51\arcsec\ length, providing a spectral resolution of
$\Delta\lambda\sim12$~\AA\ and ample real estate for sky subtraction.  Basic
properties of the targets and a summary of continuum polarization values
measured at each epoch are listed in Tables~1 and 2, respectively.  Only
PKS~0405$-$123 showed such weak polarization in the initial observation that
accurate spectropolarimetry was impractical\footnote{A broadband measurement
made with a filter polarimeter and included in Table~2 confirms that this low
polarization persisted as late as 1999 Oct.}. With that lone exception, the
new spectropolarimetric results are broadly consistent with ``white-light''
and filtered measurements from the 1970's and 1980's (e.g., Stockman, Moore,
\& Angel 1984; Berriman et al. 1990; Webb et al. 1993). In the
spectropolarimetric results, the maximum position angle difference between the
optical polarization and radio axis on either arcsecond (as) or milliarcsecond
(mas) scales is $|\Delta\theta|=28\arcdeg$ (Table~3), so the new observations
substantiate a long-term relationship between the two quantities.

For the 7 objects observed in detail, polarimetric variability was small, when
present at all (see \S4.2).  Thus, the data for each object were coadded,
weighting according to the statistical quality of each observation. The
combined results are shown in Figures~$1-7$, plotted as observed quantities
{\it vs.\/} rest wavelength. Each figure portrays, from bottom to top, the
total flux spectrum $F_\lambda$, the Stokes flux $q^\prime\times F_\lambda$
for a coordinate system aligned with the overall position angle of
polarization, the rotated Stokes parameter $q^\prime$ in percent, and the
electric-vector position angle $\theta$.  For objects whose polarization
position angles do not vary strongly with wavelength, such presentations are
preferable to plots of the degree of polarization $P$ and polarized flux
$P\times F_\lambda$ because they avoid having to apply statistical corrections
for the positive bias of measurement errors on $P$.  The final polarimetric
signal-to-noise ratio for each object is $q^\prime/\sigma_{q^\prime}=6-10$ per
20~\AA\ wide bin, which is somewhat broader than a spectral resolution element.

The sample was chosen primarily to be tractable with a 2.3~m telescope.
Generally, sources are optically bright, $V<17$, and modestly polarized,
$P\gtrsim1\%$. All lie above a galactic latitude of 40\arcdeg\ except 4C~34.47
($b=32\arcdeg$).  In any case, the lack of measurable polarization in the
emission lines in any object (\S3) rules out a significant interstellar
component due to aligned grains in either the Milky Way or quasar host galaxy.
The targets have total (core + lobes) radio luminosities $26 \lesssim
\log(L_\nu (\rm 5~GHz)) \lesssim 27$ (W~Hz$^{-1}$) and core 5~GHz-to-optical
flux ratios greater than 10 (Table~1).  One of the most core-dominant of the
radio sources, 4C~34.47, is also a known superluminal source (Barthel et al.
1989). Redshifts span the range $0.20<z<0.62$, so the \hbeta+[\ion{O}{3}]
complex lies in the observed spectral range for all objects; lower-$z$ objects
also provide \halpha, and 3C~95 exhibits \ion{Mg}{2} $\lambda2800$.  In
general, the emission-line components are very prominent in total flux.

Because targets were selected from a list already known to be polarized and
with an orientation related to the radio structure, our sample is not
well-suited in isolation for statistical evaluation of polarization properties
{\it vs.\/} the predictions of a particular model (e.g., the Unified Scheme).
Instead, our goal is to examine what can be learned about the polarizing
mechanism in the aligned objects and possibly about the central regions of
radio-loud quasars. Comparisons with spectropolarimetric compilations of other
AGN may be informative but relevant selection criteria should be kept in
mind.  For example, Cohen et al. (1999) presented observations of 13 low-$z$
radio galaxies.  That sample contains 8 narrow-line objects whose nuclei are
presumably highly obscured; 6 of these reveal broad lines in polarized flux.
Of the 5 broad-line radio galaxies, one was chosen by Cohen et al. because
other observations suggested that it was heavily obscured. The remainder show
no tendency toward alignment between optical polarization and radio structure,
some contain large starlight components in their optical continua, and only
one exhibits a true quasar-like spectrum.  More recently, Malkan (2000) has
made a study of several low-polarization QSOs of a variety of types, including
3 radio-loud objects.

A few of the targets observed here have been studied with spectropolarimetric
techniques by others.  Antonucci (1988) presented data of Miller and Goodrich
on 4C~34.47 and Ton~202, PKS~0405$-$123 was studied by Antonucci et al. (1996),
and Malkan (2000) discusses 4C~09.72.  In none of these results do the
emission lines or 3000~\AA\ bump appear in polarized light, though the
previous data on 3C~34.47 and Ton~202 are of low signal-to-noise ratio and the
polarization of PKS~0405$-$123 was very weak also when studied by Antonucci
($1990-91$; $P\approx0.04$\%).

\section{Spectra in Polarized Light}

From inspection of Figures $1-7$, it is clear that the following
characteristics are universal among the objects studied:

\begin{enumerate}

\item a position angle of polarization that is essentially constant with
wavelength,

\item a degree of
polarization\footnote{More accurately, the rotated Stokes parameter $q^\prime$
(\%)} that is diminished at the locations of broad and narrow emission lines,
and

\item a reduction in polarization in the interval occupied
by the 3000~\AA\ bump: 2500~\AA\ $\lesssim \lambda_{\rm em} \lesssim$ 5000~\AA.

\end{enumerate}

Malkan (2000) notes that the combination of all 3 of the above trends is found
{\it only\/} in the radio-loud objects he studied.  The results contradict a
claim of Stockman et al. (1984) that the degree of polarization in a broad
blue filtered bandpass generally exceeds that measured in the red. However,
Stockman et al. point out that definite wavelength dependence was detected in
only 2 of their objects.  One is the $z=2.23$, radio-quiet Broad Absorption
Line QSO 1246$-$057 (e.g., Schmidt \& Hines 1998), where the optical band
examines a region well short of the 3000~\AA\ bump.  Evidence among the
remainder of their sample is largely statistical in nature owing to large
measurement uncertainties.  Of the 4 objects which are in common with our
spectropolarimetric sample, only Ton 202 showed a blue-red polarization
difference exceeding the combined 1$\sigma$ error bar: in that case,
$P_B<P_R$, as we find.

Trends 2) and 3) above produce a spectrum of Stokes flux $q^\prime\times
F_\lambda$ which increases smoothly toward shorter wavelengths, showing no
evidence of line emission. The confidence which can be placed on the claim of
unpolarized lines varies from object to object according to the degree of
polarization, data quality, and emission-line strength.  We have used simple
spectral synthesis techniques to estimate the fraction of the observed
broad-line flux, $f_p$, which could share the polarization characteristics of
the continuum, gauging the appearance of a polarized emission line in the
models by eye. Because of the difficulty of setting a ``continuum'' level
beneath the 3000~\AA\ bump, we confine this test to the most prominent broad
emission line in the spectrum: \halpha, \hbeta, or \ion{Mg}{2} $\lambda2800$,
as appropriate. The results are listed in Table~4 as 3$\sigma$ upper limits to
the polarized fraction of these features.  Limits range from $f_p<0.4$ for the
\halpha\ line of 4C~34.47 to $f_p<1.5$ for the weak \hbeta\ feature in
4C~09.72.  Four of the 7 targets yield 3$\sigma$ upper limits so low that only
a portion of the line could be polarized like the continuum. This conclusion
contrasts sharply with typical polarized type 2 sources, where broad lines
generally appear {\it only\/} in polarized light, $f_p\gg1$.

It should be pointed out that the above-quoted confidence levels are quite
conservative: Several unpolarized emission features are present in most
spectra, and statistical limits for the ensemble would be tighter than for the
single brightest line. Furthermore, since the 3000~\AA\ bump is thought to
arise largely as gaseous emission within the BLR, the observed decline in
polarization within the bump in each object is strong evidence for an
unpolarized broad-line component.  As an aside, we point out that many of the
objects have sufficiently bright [\ion{O}{3}] $\lambda\lambda$4959, 5007 lines
to rule out the possibility that a significant portion of the narrow-line
emission shares the polarization of the continuum.

Our results demonstrate that the polarization of the optical/radio aligned
quasars studied here is a property solely of the continuum which underlies
3000~\AA\ bump and is not shared by the emission lines. We have characterized
the shape of this smoothly rising polarized continuum by a best-fit power law
in Stokes flux ($q^\prime\times F_\nu)\propto\nu^\alpha$.  The resulting
spectral indices $\alpha$ are listed in Table~4.

\section{The Polarizing Mechanism and Structure of the Nucleus}

The results presented above lead to several general conclusions regarding the
polarizing mechanism in these radio-optical aligned quasars:

\begin{enumerate}

\item Each object is dominated by a single polarizing mechanism.

\item This mechanism acts entirely, or in large part, on
the continuum of the object.  If scattering is involved, the particles cannot
lie substantially outside either of the emission-line regions. Electrons are
possible scatterers, but considering that the BLR in objects of this
luminosity may be as large as 1~pc in radius (Kaspi et al. 2000), dust
particles may also be viable if the grains are entrained in a wind near the
condensation radius (Kartje 1995).

\item The fact that the polarizing mechanism is aware of the radio axis
on pc through Mpc scales implicates an origin that is related to the
collimation of, or radiation by, the particle beam: i.e. the accretion disk or
emission from the jet itself. These two alternatives are discussed in detail
below.

\end{enumerate}

\subsection{A Scattering Disk?}

It is popular to interpret the ``big blue bump'' which dominates the
optical/UV portion of the spectral energy distribution of a typical AGN as
emission from an accretion disk that surrounds the compact object. Though the
shape of this bump is not a power-law over long spectral baselines, Francis et
al. (1991) measured optical-near UV continuum slopes for a large number of
radio-loud and -quiet QSOs, after correcting for the contributions of the
Balmer-continuum and \ion{Fe}{2} emission.  The resulting histogram reveals
that the great bulk of the objects falls in the range $-1.0 < \alpha < +0.3$,
where $\alpha$ is the spectral index $F_\nu\propto\nu^\alpha$. It can be seen
from comparison with Table~4 that this range also encompasses all of the slopes
measured for the {\it polarized flux\/} spectra of the radio-optical aligned
quasars studied here. The continuum shapes of these objects are therefore
consistent with a spectrally-neutral polarizing mechanism applied to the light
of the big blue bump.

Unfortunately, actual models of the polarization produced by physically-thin
accretion disks are not as categoric (Koratkar \& Blaes 1999; Koratkar et al.
1995; Antonucci et al. 1996). Briefly, simple optically-thick disk models
imply a high opacity due to electron scattering, hence a substantial
polarization when viewed at sufficiently large inclination, $i \gtrsim
45\arcdeg$.  This prediction is modified in spectral regions where the
absorption opacity is also high, so strong polarimetric signatures are
expected near bound-free edges. Spectropolarimetric observations have been made
of a number of polarized QSOs in the region around the Lyman edge (e.g.
Koratkar et al. 1995, 1998; Antonucci et al. 1996; Schmidt \& Hines 1998). The
data fail to show either the strong predicted continuum polarizations or the
expected discontinuities.  Moreover, scattering in an optically-thick,
physically-thin disk results in a polarization $E$-vector which is {\it
perpendicular\/} to the disk axis; the latter is thought to approximately
coincide with the radio axis.  This relationship is, of course, opposite to
the trend observed among the radio-optical aligned quasars. Thus, simple thin
disks seem to be ruled out on a number of accounts, a conclusion which seemed
inevitable even in 1986 (Antonucci 1988). Suggested remedies to the situation
include physically thick disks, winds, and additional scattering structures
(see, e.g., Kartje 1995).  Unfortunately, such complications reduce the
predictability of models and as a result the utility of polarimetry as a
diagnostic of the central structure of the nucleus.

A discussion of scattering disks in AGN would be incomplete without mentioning
the case of OI~287.  This object is an unusually strong narrow-line
source\footnote{Indeed, Antonucci et al. (1993) term it a Quasar 2.} whose
polarization is extremely high ($P\sim8\%$), constant over time, and shows a
position angle that is precisely aligned with the double-lobed radio structure
(Rudy \& Schmidt 1988). Like the radio-optical aligned quasars, the narrow
emission lines are unpolarized.  However, in OI~287 broad \hbeta\ and \hgamma\
are evident in polarized light (Goodrich \& Miller 1988; see also Antonucci,
Kinney, \& Hurt 1993), a fact which motivated a model involving obscuration
and scattering of light originating in the central source and BLR by a
physically (and optically?) thin disk that is viewed very nearly edge-on.
While OI~287 demonstrates the feasibility of thin scattering disks resulting
in strong parallel polarizations, its overwhelming narrow-line spectrum and
polarized broad lines indicate that the source is viewed at a high inclination
and may well be the product of a polarizing mechanism unrelated to that
dominating the objects studied here.

\subsection{Synchrotron Alternatives}

Synchrotron emission is a well-known source of optical continuum and
polarization in certain classes of AGN. It originates both from resolved,
kpc-scale jet features such as those which emanate from several radio-loud AGN,
and from variable components in the spectral energy distributions of blazars.
The latter presumably also arise in small-scale jet structures which in the
Unified Scheme (e.g. Antonucci 1993) are Doppler-boosted when a relativistic
particle beam is viewed from near the axis of ejection.  Recall that in the
Unified Scheme, broad-line quasars are seen from a sufficiently small
inclination that the torus does not intercept sightlines to the nucleus
($i\lesssim45\arcdeg$), so a weak optical synchrotron component from the
relativistic jet would not be a great surprise.

With the exception of the enormous jet of 3C~273, the resolved optical jets
associated with local AGN have projected lengths of a few kpc (e.g., Scarpa et
al. 1999).  At a distance typical of the sources in Table~1, they would subtend
$<$1\arcsec\ on the sky and be very difficult to discern against the bright
nuclear source. Both well-studied local examples -- M87 and 3C~273 -- show
superluminal motion (Biretta \& Meisenheimer 1993; Unwin et al. 1985 and
references therein) and have been mapped at $\lesssim$1\arcsec\ resolution in
spectral index and polarization (Perlman et al. 1999; Conway \& R\"oser 1993).
The optical jets are not only associated with radio counterparts but are
aligned with radio structure on the mas scale as well as with the giant lobes.
The jets of both M87 and 3C~273 exhibit optical $E$-vectors of polarization
that are generally perpendicular to the projected jet axes in the inner
regions ($B$-fields parallel to the jets), flipping by $\sim$90\arcdeg\ in
knots that are presumed to be the sites of shocks. As a result, the
polarization of each jet summed over its entire length is modest
($P\lesssim10$\%).  Finally, optical spectral indices measured for the local
jet sources are: M87, $\alpha\approx-0.65$ independent of location (Biretta \&
Meisenheimer 1993) and 3C~273, $-1.7<\alpha<-0.8$ (Conway \& R\"oser 1993).
Radio-optical spectral indices of jets associated with PKS~0521$-$365, 3C~371,
and PKS~2201+044 measure $\alpha_{\rm RO}\approx-0.8$ (Scarpa et al. 1999).
These slopes compare well with the spectral indices of polarized flux for the
targets studied here ($-1.1\lesssim\alpha\lesssim-0.8$), except for 4C~34.47
($\alpha\approx+0.1$; Figure~6 and Table~4).

The difficulty with the hypothesis that kpc-scale jets are the origin of
polarization in the radio-optical aligned quasars is the low optical
luminosities of known examples. Monochromatic values at $V$ summed over the
lengths of the above-mentioned jets range from $L_{5500} \sim 6\times10^{35}$
ergs s$^{-1}$ \AA$^{-1}$ for PKS~2201+044 to $\sim$$7\times10^{38}$ ergs
s$^{-1}$ \AA$^{-1}$ for 3C~273\footnote{Cosmological parameters
$H_0=75$~km~s$^{-1}$~Mpc$^{-1}$ and $q_0={1\over2}$ are assumed.}.  These must
be compared with the optical luminosity for a typical quasar from our sample
($V\approx16$, $z\approx0.35$) of $\sim$10$^{42}$ ergs s$^{-1}$ \AA$^{-1}$.
Indeed, if the jet of 3C~273 were unresolved from the central source, it would
give rise to a net optical polarization for the quasar of $P<0.01$\%. Of
course, it is possible that our selection of the most highly polarized quasars
has also identified sources with unusually bright or polarized optical jets.
However, the structures would have to be at least an order of magnitude more
luminous than any local example and be uniformly polarized near the synchrotron
maximum of $P\approx70$\% to yield the levels of polarization listed in
Table~2.

More promising is the possibility of synchrotron emission from a misdirected
blazar core component or an intrinsically weak end-on blazar. Except for
4C~34.47, the spectral indices of polarized flux of the radio-optical aligned
quasars are consistent with the optical-near UV range measured for blazars
(e.g. Landau et al. 1986), and a synchrotron component with moderately strong
polarization ($P\sim10\%$) that provides $\lesssim$10\% of the total continuum
flux would account for the observed levels of polarization.  The canonical
benchmark for a highly-polarized quasar has been that the broadband optical
polarization exceed $P=3\%$ (e.g. Moore \& Stockman 1981; Stockman et al.
1984; Wills et al. 1992). This criterion was based on the bimodal distribution
of polarization observed among survey QSOs (Stockman 1978), a fact that
suggested different polarizing mechanisms dominate in the two regimes.  While
it is true that quasars exceeding the 3\% limit are very likely to show
blazar-like properties, it is quite possible -- indeed probable -- that some
weakly-polarized objects also contain feeble synchrotron components.
Remarkably, the best example of a miniblazar is found in the prototypical
emission-line quasar 3C~273 (Impey, Malkan, \& Tapia 1989; Wills 1989). Highly
variable in polarization but generally showing $P<0.5$\% at optical
wavelengths, 3C~273 occasionally exhibits ``flares'' when the $I$-band
polarization reaches 2\% (Courvoisier et al. 1988; Impey et al. 1989; de Diego
et al. 1992).  Dilution by the 3000~\AA\ and big blue bumps cause the degree
of polarization to decline toward the blue and the emission lines are measured
to be patently unpolarized (Smith, Schmidt, \& Allen 1993; de Diego et al.
1992).  From the same observations, it was found that the spectrum of
polarized flux in the optical-UV is rather steep, $\alpha<-1.5$, and variable.
Though the polarization position angle has been measured to vary through a
range of at least 120\arcdeg, values tend to cluster around a mean of
$\theta\sim60\arcdeg$ (Impey et al. 1989), close to the $\theta=43\arcdeg$\
orientation of the resolved jet and radio lobes. Thus, 3C~273 could well be
classified as a quasar that shows alignment between its optical polarization
and radio structure.  A lower-luminosity analog is the type-1 Seyfert galaxy
NGC~4151, which contains a variable continuum component that has been studied
extensively throughout the electromagnetic spectrum and whose optical
polarization lies within $\sim$10\arcdeg\ of the axis of the two-sided
arcsecond-scale radio jet (e.g., Schmidt \& Miller 1980; Pedlar et al. 1993).

The classic test for blazar activity is variability. The polarimetric data
available to us (Stockman et al. 1984; Webb et al. 1993) coupled with the
observations summarized in Table~2 provide 5$-$10 epochs of broadband
measurements for each object in our sample.  The data span a total time base
of typically 20~yr and sample intervals of about a month to a decade.  A
simple $\chi^2$ analysis of the measurements uncovers a few interesting
cases.  Foremost is the aforementioned PKS 0405$-$12, which was found to have
$P\ge0.50$\% on two occasions by Stockman et al. (1984) but has always been
measured at $<$0.30\% since. The position angle of polarization, which is
probably a more robust indicator since it is not as sensitive to differences
in waveband or to variable dilution, shows values for this quasar that range
over nearly the entire compass.  Even in 1978$-$80, when the source was
comparatively strongly polarized, a range of $\Delta\theta=24\arcdeg$\ was
measured. Unfortunately, PKS 0405$-$12 has been so weakly polarized of late
that accurate spectropolarimetry has been impossible (see also Antonucci et
al. 1996).

Three other targets from our sample show some evidence for polarimetric
variability.  The $\chi^2$ index for the position angle measurements of
Ton~202, 4C~37.43, and 4C~09.72 are 35, 23, 23 for 9, 4, 6 degrees of freedom,
respectively, corresponding to probabilities of chance occurrence of
$10^{-4}-10^{-3}$.  The range of variation is not great, amounting to
$\Delta\theta\sim30$\arcdeg\ for 4C~09.72 and 4C~37.43 and $\sim$20\arcdeg\
for Ton~202.  However, these cases are unlikely to be statistical flukes,
since measurements for B2~1208+32 and 4C~34.47, acquired to similar accuracy
with the same instruments and over similar baselines, cover a total span in
$\theta$ of just 12\arcdeg\ and 8\arcdeg, respectively, and yield
$\chi^2_\theta=3$ for 6 degrees of freedom for each object. Unfortunately, the
detection of variability requires high relative precision in the measurements,
and available data for these quasars are not of uniformly high quality. If
3C~273 were not virtually on our doorstep, it is likely that its miniblazar
source would remain undiscovered even today.

While polarimetric variability may be a sufficient condition for identifying
the synchrotron mechanism, it may not be necessary. Even the most active
blazars often exhibit ``preferred'' polarization position angles in extensive
monitoring campaigns (e.g. Angel et al. 1978), suggesting a systemic axis
possibly associated with that of the accretion disk. The synchrotron
components proposed here would be much more stable than those of classical
blazars, but this is qualitatively consistent with jets viewed at larger
angles to the ejection axis. A strong link between the existence of a beamed
optical component and core-dominant radio structure is well-established (e.g.
Impey \& Tapia 1988; Wills et al. 1992), and strong optical polarization is
almost always seen in superluminal sources.  It is therefore interesting that
the most core-dominant of our targets, 4C~34.47, with $\log (F_{\rm
core}/F_{\rm tot})\approx-0.3$, lies in the range where a majority of sources
would show high polarization (Impey et al. 1989). 4C~34.47 has also shown
superluminal motion with the rather modest apparent speed of $\beta_{\rm
app}\sim3$ (Barthel et al. 1989).  The projected expansion axis was oriented
at $\theta=167\arcdeg$, very nearly aligned with the large-scale structure and
only 19\arcdeg\ from the optical polarization position angle.  A simple
beaming model suggests that the jet lies within 44\arcdeg\ of the line of
sight.  Once again we draw a parallel to 3C~273, which is a superluminal
source showing $5\lesssim\beta_{\rm app}\lesssim8$ and a corresponding viewing
angle of $<$20\arcdeg\ (Unwin et al. 1985).  On the other hand, the unusually
flat spectral index of polarized flux measured for 4C~34.47 would require a
miniblazar source of unprecedented properties -- one with a remarkably flat
synchrotron spectrum and/or strong intrinsic wavelength dependence to its
polarization.

\section{Discussion}

The preceding sections present modest evidence that the optical polarization
of radio quasars may be due largely to synchrotron emission from misdirected
or intrinsically weak blazar core components. Because this hypothesis implies
the presence of an exposed relativistic jet, and such jets are presumably
absent in radio-quiet objects, it might be expected that radio quasars would
be over-represented among AGN with small but significant polarization in
large-scale surveys.  Of course, this implies equality between the classes
with respect to other variables (dust content, orientation, starlight
fraction, etc.). In their statistical analysis of the polarization of
Palomar-Green (PG) QSOs, Berriman et al. (1990) discriminated explicitly
according to radio properties.  Their Figure~3 indeed shows that a
significantly larger fraction (5:17) of radio-loud objects (those with $F_\nu
(\rm 5~GHz)/F_\nu (\rm opt) > 25$) are found with $P>1$\% than for the
radio-quiet sample (9:80).  Four of the radio-luminous PG sources are
spectropolarimetric targets studied here. However, Berriman et al.'s breakdown
according to radio core dominance showed no correlation within either the
radio-loud or -quiet samples, and a major conclusion of that paper was the
general lack of evidence for synchrotron components in the sources studied.
The techniques applied by Berriman et al. (1990) were effective at
discriminating against highly polarized AGN of either the scattering or blazar
variety, but since the PG catalog is derived from optical selection criteria,
the radio-loud subsample is small. The criteria used -- polarimetric
variability and the dependence of $P$(\%) on radio core dominance and optical
color -- were inefficient at identifying weak nonthermal components when the
observable itself was known to a typical accuracy of only $\sigma_P\sim0.2$\%
($P/\sigma_P\lesssim5$). Indeed, the prototype miniblazar, 3C~273, fails to
stand out in any of the diagnostic tests made by Berriman et al., despite the
fact that it is a particularly bright and easily measured member of the PG
catalog.  The weak polarization that is observed at wavelengths as long as
2~$\mu$m in 3C~273 as well as for 3 of the spectropolarimetric targets studied
here (Sitko \& Zhu 1991) indicates that the putative synchrotron components
would also be minor contributors to the observed flux in the infrared.
Unfortunately, none of the sources studied here were detected by {\it IRAS\/},
but they would make interesting {\it SIRTF\/} targets. Detailed spectral
energy distributions, plus comprehensive polarimetric monitoring and
additional investigations of the type carried out here, may be required to
confirm the existence and prevalence of synchrotron components as significant
sources to the optical continua of quasars.

\acknowledgments

Special thanks go to D. Hines, M. Malkan, and the referee, R. Antonucci, for
illuminating discussions and critical readings of the manuscript. Support for
this work was provided by NSF grants AST 91-14087 and AST 97-30792 and by the
Director of KPNO.

\clearpage

\begin{deluxetable}{lcccccc}
\tablenum{1}
\tablewidth{0pt}
\tablecaption{Object List}
\tablehead{\colhead{Quasar} & \colhead{Coord.} & \colhead{$z$} & \colhead{$V$}
& \colhead{$\log(L_\nu (\rm 5~GHz))$\tablenotemark{a}} & \colhead{$\log R\tablenotemark{b}$} & \colhead{$F_\nu (\rm 5~GHz)/F_\nu (\rm opt)$} \\
& & & & \colhead{(W~Hz$^{-1}$)} & & \colhead{(core only)}}
\startdata
3C 95         & 0349$-$146 & 0.616 & 16.2 & 26.9 & $-1.68$ & ~\,13 \\
PKS 0405$-$12 & 0405$-$123 & 0.574 & 15.8 & 27.2 & $-0.23$ & 384 \\
B2 1208+32    & 1208+322   & 0.388 & 16.0 & 25.7 & ~$\cdots$ & $\cdots$ \\
Ton 202       & 1425+267   & 0.362 & 15.7 & 25.6 & $-0.40$ & ~\,19 \\
4C 37.43      & 1512+370   & 0.371 & 15.5 & 26.0 & $-0.71$ & ~\,24 \\
3C 323.1      & 1545+210   & 0.264 & 16.7 & 26.0 & $-1.32$ & ~\,41 \\
4C 34.47      & 1721+343   & 0.206 & 16.5 & 25.8 & $+0.06$ & 353 \\
4C 09.72      & 2308+098   & 0.432 & 16.0 & 26.1 & $-0.78$ & ~\,28 \\
\enddata
\tablenotetext{a}{Total (core + lobe) luminosity computed using $H_0=75$~km~s$^{-1}$~Mpc$^{-1}$, $q_0={1\over2}$.}
\tablenotetext{b}{Core-to-lobe flux ratio at 5~GHz, from Wills \& Browne (1986).}

\end{deluxetable}

\begin{deluxetable}{llcccc}
\tablenum{2}
\tablewidth{0pt}
\tablecaption{Spectropolarimetry Log}
\tablehead{\colhead{Quasar} & \colhead{~Epoch} & \colhead{\# obs\tablenotemark{a}}
& \colhead{$P$\tablenotemark{b}} & \colhead{$\theta$\tablenotemark{b}}}
\startdata
3C 95         & 1995 Nov & ~~7 & $0.97\pm0.06$\% & $175\arcdeg\pm2\arcdeg$ \\
              & 1996 Dec & 12  & $0.73\pm0.05$\% & $177\arcdeg\pm2\arcdeg$ \\
              & 1999 Oct & ~~2 & $0.55\pm0.18$\% & ~$178\arcdeg\pm10\arcdeg$ \\ \\
PKS 0405$-$12 & 1995 Nov & ~~1 & $0.27\pm0.19$\% & ~$124\arcdeg\pm20\arcdeg$ \\
              & 1999 Oct\tablenotemark{c} & ~~1 & $0.17\pm0.14$\% & ~~$90\arcdeg\pm25\arcdeg$ \\ \\
B2 1208+32    & 1996 Mar & ~~6 & $1.14\pm0.08$\% & ~~$18\arcdeg\pm2\arcdeg$ \\
              & 1996 Jun & ~~1 & $1.32\pm0.25$\% & ~~$14\arcdeg\pm6\arcdeg$ \\
              & 1996 Dec & ~~6 & $1.05\pm0.05$\% & ~~$17\arcdeg\pm2\arcdeg$ \\
              & 1997 Dec & ~~2 & $0.65\pm0.27$\% & ~~~$15\arcdeg\pm12\arcdeg$ \\ \\
Ton 202       & 1996 Mar & ~~2 & $1.62\pm0.15$\% & ~~$67\arcdeg\pm3\arcdeg$ \\
              & 1996 Jun & ~~4 & $1.85\pm0.12$\% & ~~$64\arcdeg\pm2\arcdeg$ \\
              & 1997 Apr & ~~2 & $1.40\pm0.16$\% & ~~$69\arcdeg\pm3\arcdeg$ \\
              & 1997 Dec & ~~2 & $1.88\pm0.26$\% & ~~$75\arcdeg\pm4\arcdeg$ \\
              & 1998 Apr & ~~3 & $1.67\pm0.18$\% & ~~$69\arcdeg\pm3\arcdeg$ \\
              & 1999 May & ~~2 & $1.97\pm0.09$\% & ~~$74\arcdeg\pm2\arcdeg$ \\ \\
4C 37.43      & 1996 May & ~~7 & $1.24\pm0.04$\% & $101\arcdeg\pm2\arcdeg$ \\
              & 1999 May & ~~3 & $1.26\pm0.09$\% & $100\arcdeg\pm2\arcdeg$ \\ \\
3C 323.1      & 1996 Jun & ~~5 & $1.35\pm0.04$\% & ~~$23\arcdeg\pm2\arcdeg$ \\
              & 1998 Apr & ~~5 & $1.72\pm0.14$\% & ~~$20\arcdeg\pm3\arcdeg$ \\ \\
4C 34.47      & 1996 Jun & ~~2 & $0.94\pm0.09$\% & $146\arcdeg\pm3\arcdeg$ \\
              & 1998 May & ~~1 & $1.13\pm0.15$\% & $146\arcdeg\pm4\arcdeg$ \\
              & 1998 Sep & ~~3 & $0.99\pm0.10$\% & $149\arcdeg\pm3\arcdeg$ \\
              & 1998 Nov & ~~1 & $0.87\pm0.15$\% & $141\arcdeg\pm5\arcdeg$ \\
              & 1999 May & ~~1 & $0.89\pm0.09$\% & $149\arcdeg\pm3\arcdeg$ \\ \\
4C 09.72      & 1996 Dec & 12  & $1.32\pm0.04$\% & $118\arcdeg\pm2\arcdeg$ \\
              & 1997 Dec & ~~3 & $1.36\pm0.11$\% & $112\arcdeg\pm3\arcdeg$ \\
              & 1998 Nov & ~~1 & $1.23\pm0.19$\% & $117\arcdeg\pm5\arcdeg$ \\
\enddata
\tablenotetext{a}{Number of observations.}
\tablenotetext{b}{Summed over a broad continuum interval near the middle
of the observed spectrum.}
\tablenotetext{c}{``White-light'' ($\lambda\lambda3200-8600$) measurement
obtained with a filter polarimeter.}
\end{deluxetable}

\begin{deluxetable}{lcclc}
\tablenum{3}
\tablewidth{0pt}
\tablecaption{Optical Polarization and Radio Structure}
\tablehead{\colhead{Quasar} & \colhead{$P_{\rm opt}$}
& \colhead{$\theta_{\rm opt}$} & \colhead{Radio Axis\tablenotemark{1}} & $\theta_{\rm opt} - \theta_{\rm rad}$ \\
& \colhead{(\%)} & & \colhead{(as; mas)}}
\startdata
3C 95         & 0.82 & 176\arcdeg & 164\arcdeg; $\cdots$      & 12\arcdeg \\
PKS 0405$-$12 & 0.27 & 124\arcdeg & ~\,12\arcdeg; 170\arcdeg: & $-$68\arcdeg~~ \\
B2 1208+32    & 1.04 & ~18\arcdeg & ~~~3\arcdeg; $\cdots$     & 15\arcdeg \\
Ton 202       & 1.74 & ~68\arcdeg & ~\,53\arcdeg; $\cdots$    & 15\arcdeg \\
4C 37.43      & 1.24 & 100\arcdeg & 109\arcdeg; $\cdots$      & $-$9\arcdeg~ \\
3C 323.1      & 1.47 & ~22\arcdeg & ~\,20\arcdeg; $\cdots$    & ~2\arcdeg \\
4C 34.47      & 0.95 & 148\arcdeg & 162\arcdeg; 167\arcdeg    & $-$14\arcdeg~~ \\
4C 09.72      & 1.32 & 117\arcdeg & 145\arcdeg; $\cdots$      & $-$28\arcdeg~~ \\
\enddata
\tablenotetext{1}{References for radio structure --- 3C~95: Price et al. 1993;
PKS~0405$-$12: Fanti et al. 1977, Rusk \& Seaquist 1985;
B2~1208+32: Fanti et al. 1977; Ton~202, 4C~37.43, 3C~323.1, \& 4C~09.72: Kellerman et al. 1994;
4C~34.47: Barthel et al. 1989}
\end{deluxetable}

\begin{deluxetable}{lcc}
\tablenum{4} \tablewidth{0pt} \tablecaption{Characteristics of the Polarized
Spectra} \tablehead{\colhead{Quasar} &
\colhead{$\alpha$\tablenotemark{a}~($\pm$$\sim$0.2)} &
\colhead{$f_p$\tablenotemark{b}}} \startdata
3C 95         & $-$1.1 & 0.8 \\
B2 1208+32    & $-$1.0 & 1.0 \\
Ton 202       & $-$0.8 & 0.8 \\
4C 37.43      & $-$0.8 & 1.2 \\
3C 323.1      & $-$0.9 & 0.6 \\
4C 34.47      &   +0.1 & 0.4 \\
4C 09.72      & $-$0.8 & 1.5 \\
\enddata
\tablenotetext{a}{Spectral index of polarized flux $\alpha=d\log(q^\prime\times F_\nu)/d\log\nu$}
\tablenotetext{b}{3$\sigma$ upper limit to the flux fraction of the strongest broad
emission line that could share the polarization of the continuum.}
\end{deluxetable}

\clearpage

\begin{figure}
\includegraphics[bb=30 650 2 2, scale=.8, angle=0]{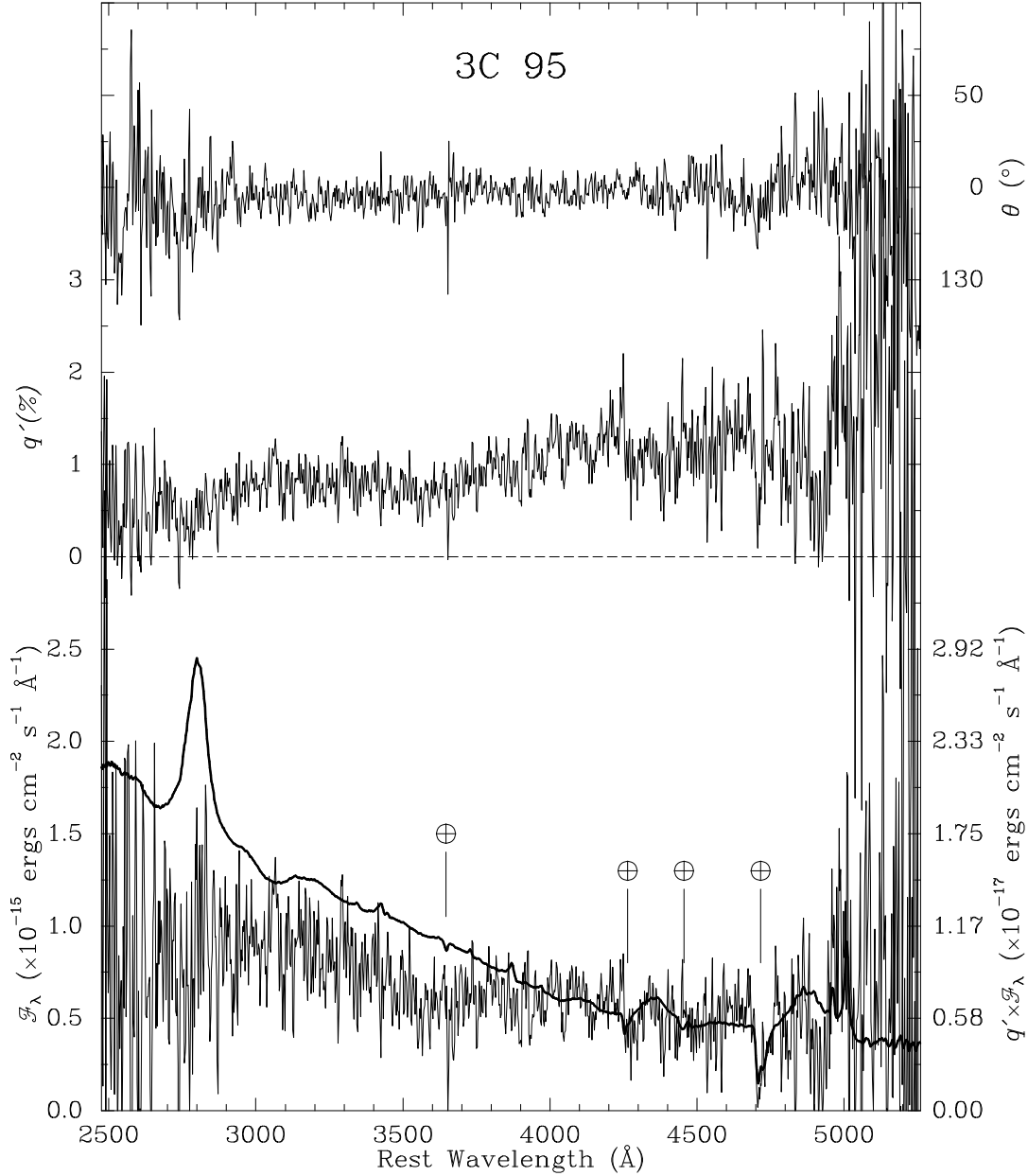}
\vskip6.2truein
\caption{Spectropolarimetry summed over all epochs for 3C~95.  {\it
Bottom panel:\/} Total flux spectrum (bold line and left ordinate) and polarized flux
$q^\prime\times F_\lambda$ (thin line and right ordinate), normalized at long wavelengths.
{\it Middle panel:\/} Stokes parameter $q^\prime$(\%) for a coordinate
system rotated to the systemic position angle of polarization.  {\it
Top panel:\/} Position angle of polarization in degrees.  Note
depressions in the degree of polarization in the ``3000~\AA\ bump'' and at each
major emission line, and the corresponding smoothly rising polarized
flux spectrum. Artifacts caused by strong terrestrial
absorption and emission features are marked.}
\end{figure}

\clearpage
\begin{figure}
\includegraphics[bb=30 620 2 2, scale=.8, angle=0]{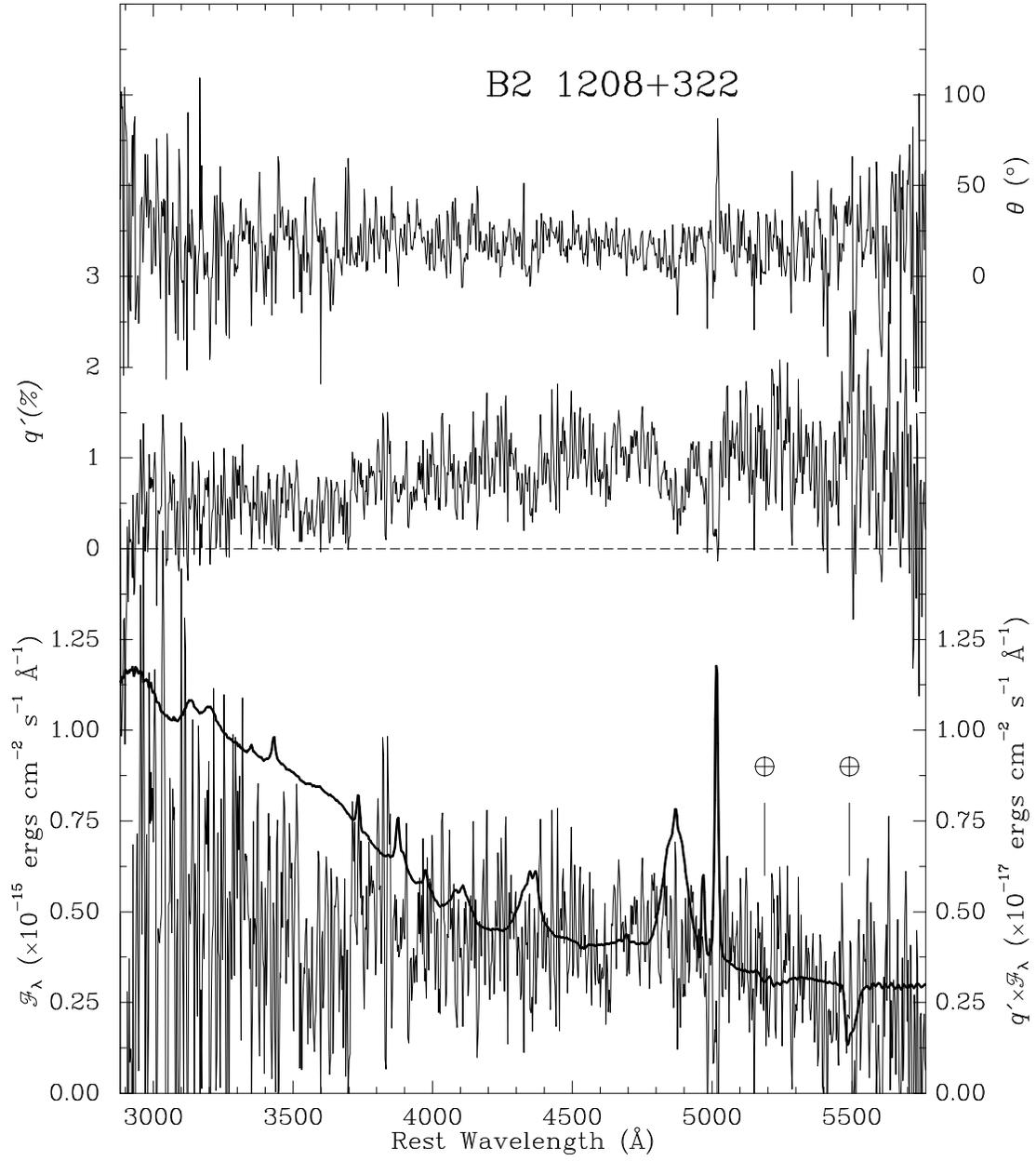}
\vskip6truein
\caption{As in Fig.~1 for B2~1208+322.}
\end{figure}

\clearpage
\begin{figure}
\includegraphics[bb=30 620 2 2, scale=.8, angle=0]{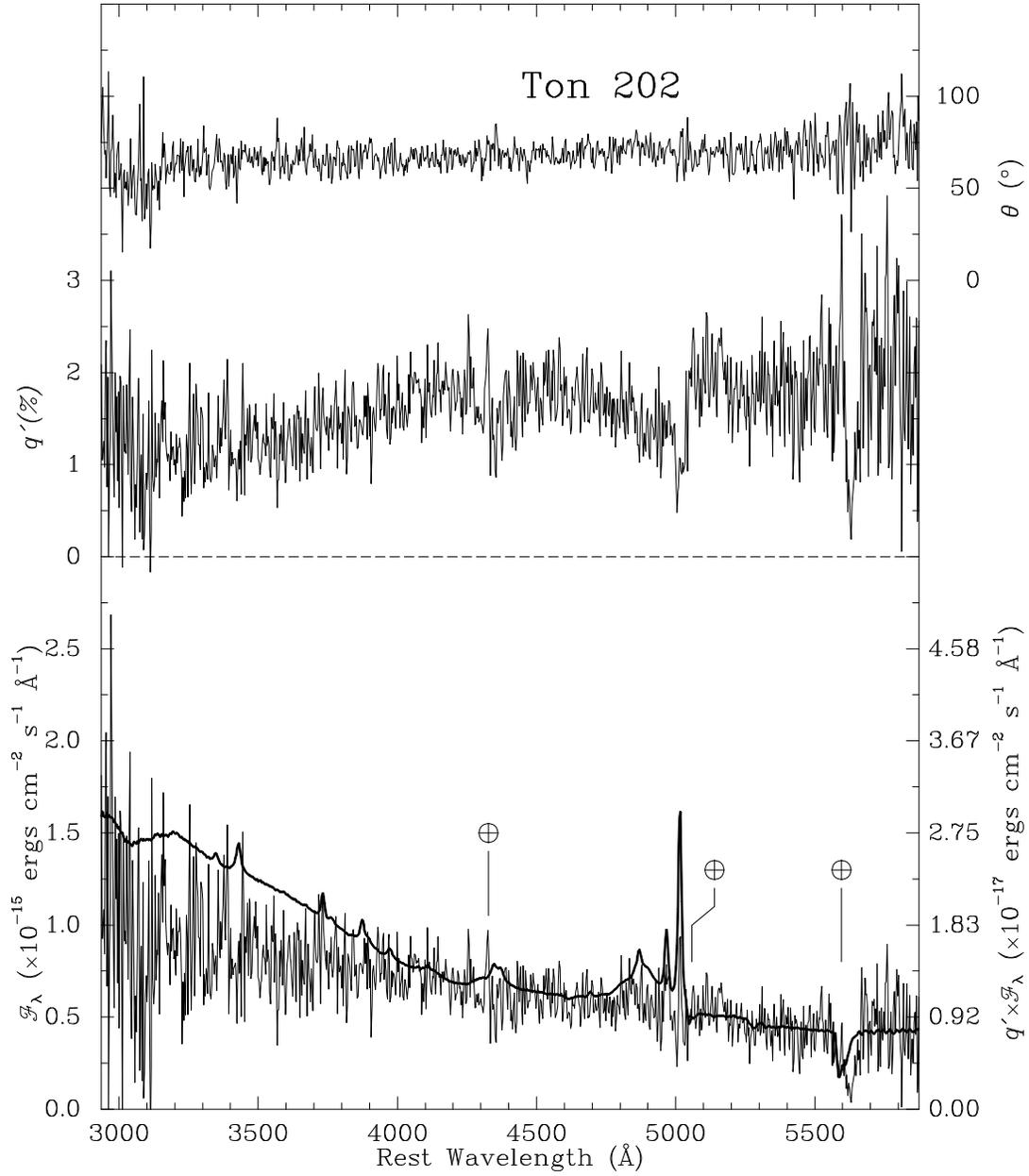}
\vskip6truein
\caption{As in Fig.~1 for Ton~202.}
\end{figure}

\clearpage
\begin{figure}
\includegraphics[bb=30 620 2 2, scale=.8, angle=0]{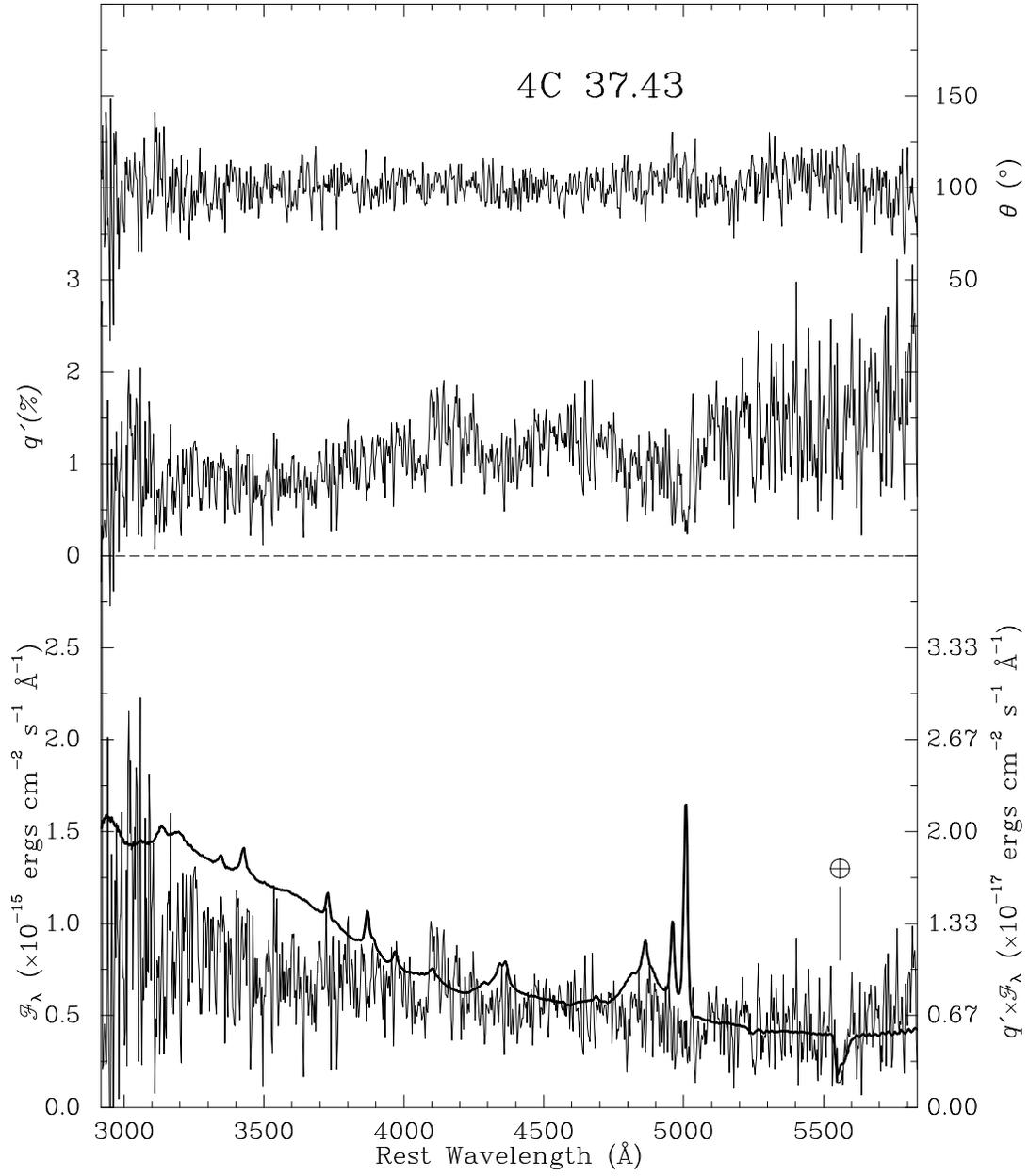}
\vskip6truein
\caption{As in Fig.~1 for 4C~37.43.}
\end{figure}

\clearpage
\begin{figure}
\includegraphics[bb=30 620 2 2, scale=.8, angle=0]{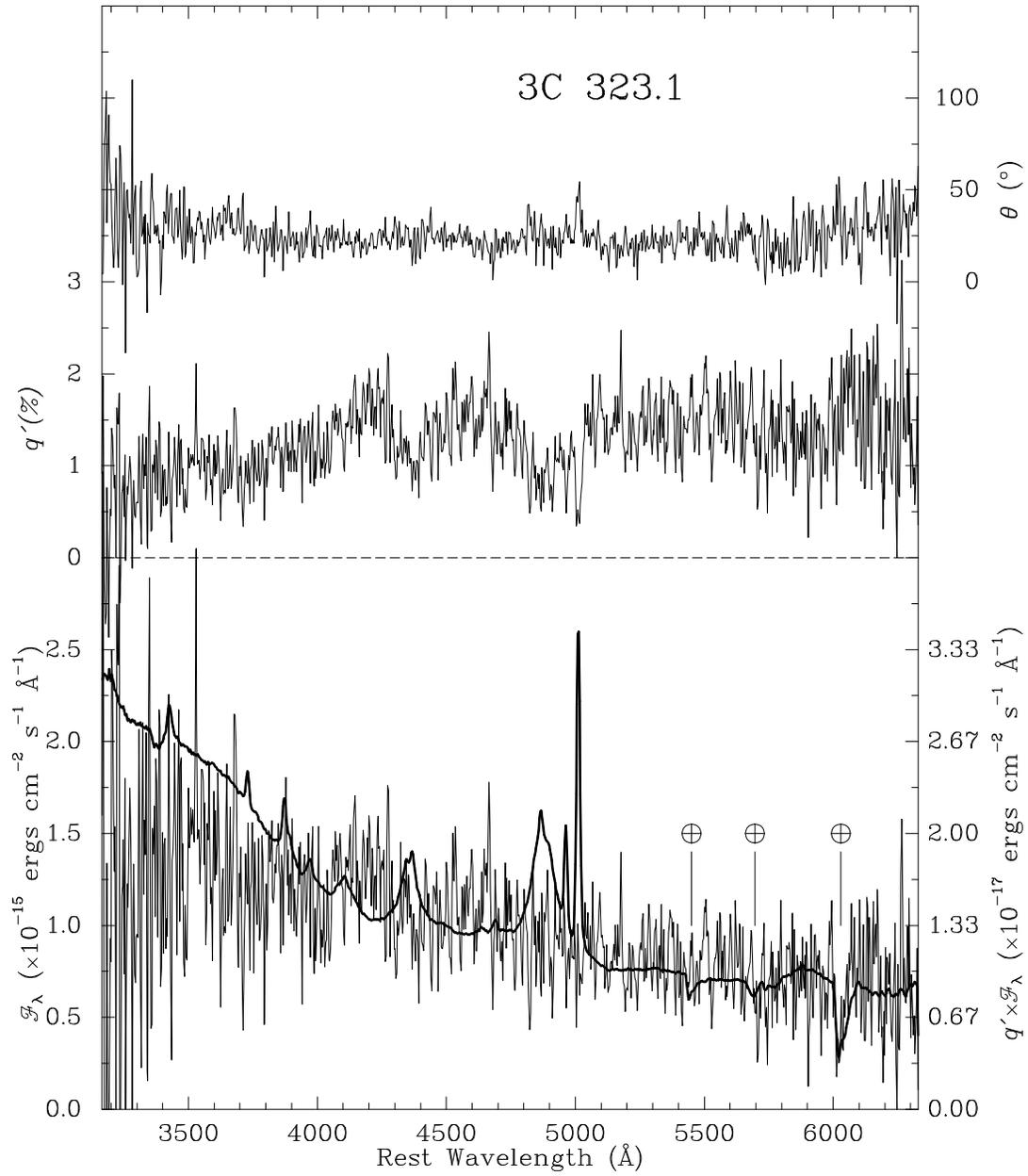}
\vskip6truein
\caption{As in Fig.~1 for 3C~323.1.}
\end{figure}

\clearpage
\begin{figure}
\includegraphics[bb=30 620 2 2, scale=.8, angle=0]{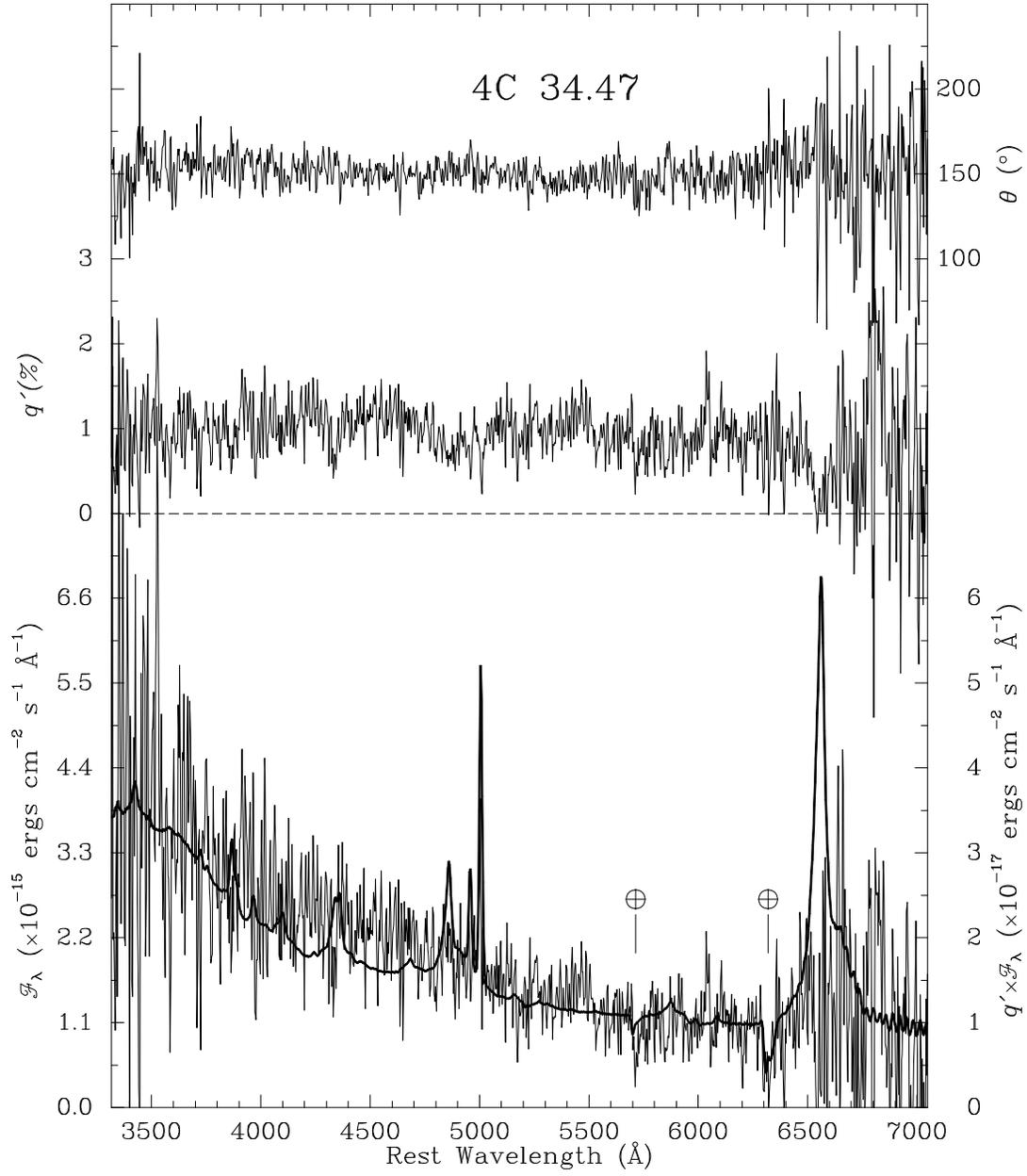}
\vskip6truein
\caption{As in Fig.~1 for 4C~34.47.}
\end{figure}

\clearpage
\begin{figure}
\includegraphics[bb=30 620 2 2, scale=.8, angle=0]{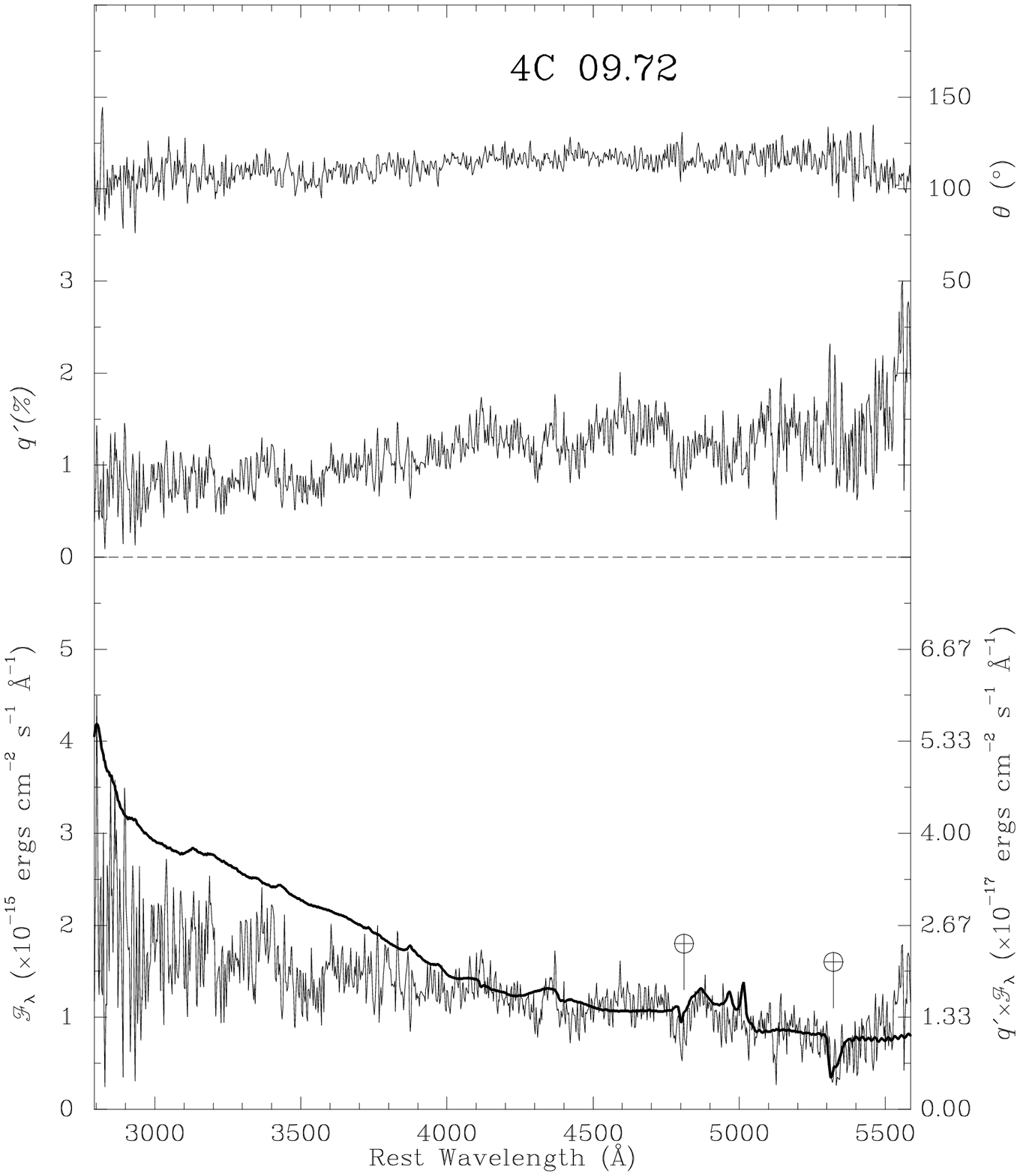}
\vskip6truein
\caption{As in Fig.~1 for 4C~09.72.}
\end{figure}


\begin{references}

\reference{} Angel, J.R.P., et al. 1978, in Pittsburgh Conference on BL Lac
Objects, ed. A.M. Wolfe (Pittsburgh: University of Pittsburgh), 117

\reference{} Antonucci, R.R.J. 1982, \nat, 299, 605

\reference{} \sameauthor\ 1983, \nat, 303, 158

\reference{} \sameauthor\ 1984, \apj, 278, 499

\reference{} \sameauthor\ 1988 in Supermassive Black Holes, ed. M.
Kafatos (Cambridge: Cambridge Univ. Press), 26

\reference{} \sameauthor\ 1993, \araa, 31, 473

\reference{} Antonucci, R., Geller, R., Goodrich, R.W., \& Miller, J.S.
1996, \apj, 472, 502

\reference{} Antonucci, R., Kinney, A.L., \& Hurt, T. 1993, \apj, 414, 506

\reference{} Antonucci, R.R.J., \& Miller, J.S. 1985, \apj, 297, 621

\reference{} Barthel, P.D., Hooimeyer, J.R., Schilizzi, R.T., Miley,
G.K., \& Preuss, E. 1989, \apj, 336, 601

\reference{} Berriman, G., Schmidt, G.D., West, S.C., \& Stockman, H.S.
1990, \apjs, 74, 869

\reference{} Biretta, J.A., \& Meisenheimer, K. 1993, in Jets in
Extragalactic Radio Sources, eds. H.-J. R\"oser \& K. Meisenheimer
(Berlin: Springer-Verlag), 159

\reference{} Capetti, A., Macchetto, F.D., \& Lattanzi, M.G. 1997,
\apj, 476, L67

\reference{} Cimatti, A., di Serego Alighieri, S., Fosbury, R.A.E.,
Salvati, M., \& Taylor, D. 1993, \mnras, 264, 421

\reference{} Cohen, M.H., Ogle, P.M., Tran, H.D., Goodrich, R.W.,
\& Miller, J.S. 1999, \aj, 118, 1963

\reference{} Conway, R.G., \& R\"oser, H.-J. 1993, in Jets in
Extragalactic Radio Sources, eds. H.-J. R\"oser \& K. Meisenheimer
(Berlin: Springer-Verlag), 199

\reference{} Courvoisier, T.J.-L., Robson, E., Hughes, D.,
Blecha, A., Bouchet, Pl, Krisciunas, K., \& Schwarz, H. 1988,
\nat, 335, 330

\reference{} de Diego, J.A., P\'erez, E., Kidger, M.R., \& Takalo, L.O.
1992, \apj, 396, L19

\reference{} Fanti, C., Fanti, R., Formiggini, L, Lari, C., \&
Padrielli, L.  1977, \aaps, 28, 351

\reference{} Francis, P., Hewett, P.C., Foltz, C.B., Chaffee, F.H.,
Weymann, R.J., \& Morris, S.L. 1991, \apj, 373, 465

\reference{} Goodrich, R.W., \& Miller, J.S. 1988, \apj, 331, 332

\reference{} Grandi, S.A. 1981, \apj, 251, 451

\reference{} Hines, D., Schmidt, G.D., Smith, P.S., Cutri, R., \& Low,
F.J. 1995, \apj, 450, L1

\reference{} Impey, C.D., Malkan, M.A., \& Tapia, S. 1989, \apj, 347, 96

\reference{} Impey, C.D., \& Tapia, S. 1988, \apj, 333, 666

\reference{} Kaspi, S., Smith, P.S., Netzer, H., Maoz, D., Jannuzi, B.T.,
\& Giveon, U. 2000, \apj, in press

\reference{} Kartje, J.F. 1995, \apj, 452, 565

\reference{} Kellerman, K.I., Sramek, R.A., Schmidt, M., Green, R.F.,
\& Shaffer, D.B.  1994, \aj, 108, 1163

\reference{} Koratkar, A., Antonucci, R.R.J., Goodrich, R.W., Bushouse, H.,
\& Kinney, A.L. 1995, \apj, 450, 501

\reference{} Koratkar, A., Antonucci, R.R.J., Goodrich, R., \& Storrs, A.
1998, \apj, 503, 599

\reference{} Koratkar, A., \& Blaes, O. 1999, \pasp, 111, 1

\reference{} Landau, R. et al. 1986, \apj, 308, 78

\reference{} Malkan, M.A. 2000, private communication

\reference{} Moore, R.L., \& Stockman, H.S. 1981, \apj, 243, 60

\reference{} Pedlar, A., Kukula, M.J., Longley, D.P.T., Muxlow, T.W.B.,
Axon, D.J., Baum, S., O'Dea, C., \& Unger, S.W. \mnras, 263, 471

\reference{} Perlman, E.S., Biretta, J.A., Zhou, F., Sparks, W.B.,
\& Macchetto, F.D. 1999, \aj, 117, 2185

\reference{} Price, R., Gower, A.C., Hutchings, J.B., Talon, S.,
Duncan, D., \& Ross, G. 1993, \apjs, 86, 365

\reference{} Rudy, R.J., \& Schmidt, G.D. 1988, \apj, 331, 325

\reference{} Rudy, R.J., Schmidt, G.D., Stockman, H.S., \& Moore, R.L.
1983, \apj, 271, 59

\reference{} Rusk, R., \& Seaquist, E.R. 1985, \aj, 90, 30

\reference{} Scarpa, R., Urry, C.M., Falomo, R., \& Treves, A.
1999, \apj, 526, 643

\reference{} Schmidt, G.D., \& Hines, D.C. 1999, \apj, 512, 125

\reference{} Schmidt, G.D., \& Miller, J.S. 1980, \apj, 240, 759

\reference{} Schmidt, G.D., Stockman, H.S., \& Smith, 1992, \apj, 398, L57

\reference{} Sitko, M.L., \& Zhu, Y. 1991, \apj, 369, 106

\reference{} Smith, P.S., Schmidt, G.D., \& Allen, R.G. 1993, \apj, 409, 604

\reference{} Stockman, H.S. 1978, in Pittsburgh Conference on BL Lac
Objects, ed. A.M. Wolfe (Pittsburgh: University of Pittsburgh), 149

\reference{} Stockman, H.S., Angel, J.R.P., \& Miley, G.K. 1979, \apj,
227, L55

\reference{} Stockman, H.S., Moore, R.L., \& Angel, J.R.P. 1984, \apj,
279, 485

\reference{} Unwin, S.C., Cohen, M.H., Biretta, J., Pearson, T.J.,
Seielstad, G.A., Walker, R.C., Simon, R.S., \& Linfield, R.P. 1985,
\apj, 289, 109

\reference{} Webb, W., Malkan, M., Schmidt, G.D., \& Impey, C.D. 1993,
\apj, 419, 494

\reference{} Wills, B.J. 1989, in BL Lac Objects, eds. L. Maraschi, T. Maccacaro,
\& M.-H. Ulrich (Berlin: Springer), 109

\reference{} Wills, B.J., \& Browne, I.W.A. 1986, \apj, 302, 56

\reference{} Wills, B.J., Netzer, H., \& Wills, D. 1985, \apj, 288, 94

\reference{} Wills, B.J., Wills, D., Breger, M., Antonucci, R.R.J., \& Barvainis, R.
1992, \apj, 398, 454

\end{references}
\end{document}